# Photon-Counting CT with Silicon Detectors: Feasibility for Pediatric Imaging

Moa Yveborg,[a] Xu Cheng,[a] Erik Fredenberg,[a] and Mats Danielsson[a]

[a]Department of Physics, Royal Institute of Technology (KTH), AlbaNova University Center, SE-106 91 Stockholm, Sweden;

## ABSTRACT

X-ray detectors made of crystalline silicon have several advantages including low dark currents, fast charge collection and high energy resolution. For high-energy x-rays, however, silicon suffers from its low atomic number, which might result in low detection efficiency, as well as low energy and spatial resolution due to Compton scattering. We have used a monte-carlo model to investigate the feasibility of a detector for pediatric CT with 30 to 40 mm of silicon using x-ray spectra ranging from 80 to 140 kVp. A detection efficiency of 0.74 was found at 80 kVp, provided the noise threshold could be set low. Scattered photons were efficiently blocked by a thin metal shielding between the detector units, and Compton scattering in the detector could be well separated from photo absorption at 80 kVp. Hence, the detector is feasible at low acceleration voltages, which is also suitable for pediatric imaging. We conclude that silicon detectors may be an alternative to other designs for this special case.

## 1. INTRODUCTION

The frequency of CT examinations is increasing, and CT is currently a major contributor to the total radiation dose (40-60%) from all x-ray examinations. This is particularly alarming considering the fact that CT scans only amount to between five and eight percent of the total number of x-ray examinations, and there is a growing concern about radiation induced cancers, in particularly for children. Nevertheless, there is also a well-motivated demand for higher image quality that could provide improved information for diagnosis.[1–3]

One way of mitigating this situation would be improved instrumentation in terms of radiation sources and detectors, and significant efforts have been made in this area. In particular, several groups are developing photon-counting detectors.[4,5] Such detectors have several advantages compared to energy-integrating detectors, including energy sensitivity and elimination of electronic noise. Two applications of energy information in CT are reduction of beam hardening, and determination of the elemental composition of the object.[6] The latter is often of high diagnostic interest, e.g. to separate contrast agents from plaque and other tissue in imaging of the coronary arteries.

Most proposed photon-counting CT detectors are composed of heavy elements such as cadmium-zinc-telluride (CZT) because of its high cross section for the photoelectric effect, which results in high detection efficiency and energy resolution. Challenges include high costs and imperfect detector crystals. In this study, we investigate the feasibility of instead using detectors made of crystalline silicon for the special case of pediatric CT. Silicon has several advantages such as being easily available in large quantities, good quality and purity of the crystals, and long time proven methods for test and assembly. For x-ray detectors, this translates into reliable devices with high performance in terms of low dark currents, fast charge collection and high energy resolution.

Silicon has been successfully employed for mammography detectors in the past.[7,8] As a detector material for the higher energies encountered in CT, however, silicon suffers from its low atomic number ($Z = 14$), which might result in low detection efficiency, low energy resolution because of the relatively high cross section for Compton scattering in the detector compared to the photoelectric effect, and a degraded spatial resolution due to Compton scattering between different detector pixels. Therefore, to investigate the feasibility of photon-counting silicon detectors for CT, it is important to show that: 1) the detection efficiency is sufficiently high, 2) scattering of photons between adjacent detector pixels can be kept at a low level, and 3) photons that Compton scatter in the detector can be separated from those that are photo absorbed, so that the former do not degrade the energy resolution.

## 2. METHODS

### 2.1. Description of the detector

We propose a detector made of 500 $\mu$m thick n-doped silicon wafers, diced into 20 mm wide and 30-40 mm deep detector units, which are arranged edge-on relative to the source. Approximately 1350 units are tiled next to each other, resulting in a 700 mm overall length of the detector. Each detector unit has implanted strips of p-doped silicon, which are arranged in a fan pointing back to the radiation source to avoid parallax errors. The units are backed with a thin tungsten foil to absorb scattered radiation between detector elements.

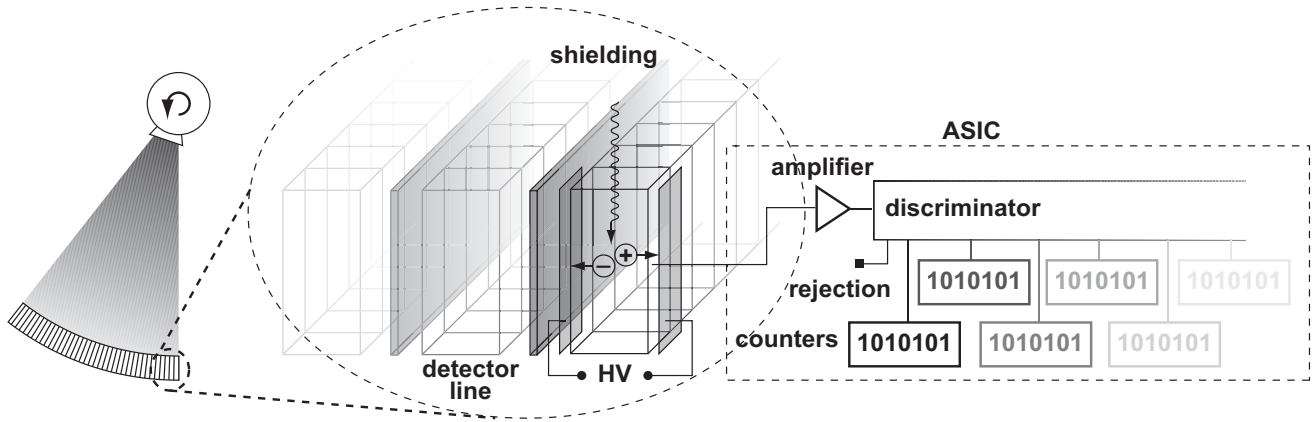

**Figure 1.** The CT detector is made of 1350 detector units that are tiled next to each other with a thin metal shielding between. Each detector unit is made of a silicon wafer and segmented into strips. The detector units and strips point back to the source.

A high voltage is applied over the detector, and charge that is released upon photon interactions in the detector drifts as electron-hole pairs towards the anode and cathode respectively. The charges may be collected with aluminum strings that are centered on the strips, but the exact implementation of these is not considered in this study and charge collection is assumed to be perfect. The readout electronics of each detector strip include a pulse height discriminator with a number of thresholds. The lowest threshold discriminates against electronic noise, and the higher ones sort the impinging photons into bins according to their energy. The number of bins is a compromise between the complexity of the readout electronics and the energy resolution, and it is not considered in this study. The low-energy noise threshold, however, affects the efficiency of the detector because some signals may be discarded as noise, and it is investigated in some detail. The intrinsic energy resolution of silicon can be approximated with a gaussian with full-width-at-half-maximum FWHM = $2.35\sqrt{E\eta\epsilon}$, where $E$ is the photon energy, $\epsilon = 3.66$ eV is the energy needed to create an electron-hole pair in silicon, and $\eta = 0.115$ is the Fano factor for silicon.[9] The FWHM ranges between 0.2 and 0.6 keV for photon energies 20-140 keV, and is small compared to charge collection errors, such as charge sharing between detector strips.

### 2.2. Monte-Carlo model of photon transport in the detector

A monte-carlo model[10] was constructed using the MATLAB software package[*]. A large number of photons are traced through the detector, and the interactions are recorded. For each photon, an energy is chosen with a probability distribution following the impinging spectrum, and a random starting position on the surface of the detector.

---

[*]The MathWorks Inc., Natick, MA

The path length ($L$) that the photon travels before an event is obtained by random sampling from the exponential probability distribution function,

$$p(L) = \mu \exp(-\mu L) \Rightarrow L = \frac{\log(\varepsilon)}{\mu(E)}, \tag{1}$$

where $\varepsilon$ is a random number between 0 and 1 drawn from a flat distribution. $\mu$ is the linear attenuation coefficient that includes all interaction processes, and the type of event is chosen randomly between (1) the photoelectric effect, (2) Compton and (3) Rayleigh scattering according to their respective cross sections.

1. If the interaction process is the photoelectric effect, the photon is fully absorbed, and all energy is detected.

2. Compton scattering is approximated with scattering on a free electron, and the polar angle ($\theta$) of the scattered photon is calculated with the Klein-Nishina differential cross section,

$$\left[\frac{d\sigma}{d\theta}\right]_{\text{KN}} = \pi r_e^2 \sin\theta \cdot \frac{1 + \cos^2\theta + k(E)^2[1-\cos\theta]^2/(1+k(E)[1-\cos\theta])}{(1+k(E)[1-\cos\theta])^2}, \tag{2}$$

where $r_e$ is the classical electron radius and $k$ is the ratio of the photon energy ($E$) and the electron energy at rest. Scattering perpendicular to the direction of travel is assumed to be uniformly distributed, and the azimuthal angle ($\varphi$) is chosen from a flat distribution. The photon energy after scattering is given by

$$E_1 = \frac{E_0}{1 + k(E_0)(1-\cos\theta)}, \tag{3}$$

where $E_0$ is the initial energy.

3. Rayleigh scattering is assumed to be fully elastic, and the Thomson cross-section is used,

$$\left[\frac{d\sigma}{d\theta}\right]_{\text{T}} = \pi \sin\theta (1 + \cos^2\theta). \tag{4}$$

$\varphi$ is again chosen from a flat distribution.

### 2.3. Simulations of the detector

A published spectrum of a tungsten target x-ray tube with a 12° anode angle,[11] filtered with 2 mm aluminum and 0.1 mm copper, was used as input to the model. Before hitting the detector, the spectrum passed a human head, consisting of 5 mm skin, 10 mm bone, 10 mm fatty tissue, and 130 mm brain matter. Published linear attenuation coefficients[12] were used to calculate the absorption.

#### 2.3.1. Detection efficiency

In principle, all photons that deposit energy in the detector can be detected, and both Compton scattered and photo absorbed photons contribute to the absorption efficiency ($N_{\text{A}}$). An approximately 300 $\mu$m thick layer at the entrance surface of the detector cannot be used for detection because of crystal imperfections caused by the dicing process. The transmission of this dicing margin ($N_{\text{DM}}$) affects the detection efficiency slightly.

A noise threshold of a few keV has to be implemented in the electronics. Such a threshold will also block the detection of some low-energy, mostly Compton scattered, quanta, and thereby reduce the detection efficiency by a factor of $N_{\text{Th}}$.

### 2.3.2. Scattering between detector elements

A photon that is scattered in the detector can meet one of four different destinies:

1. it is scattered one or several times in a detector element and absorbed within the same element or in the shielding, thus contributing to image signal;

2. it is scattered and absorbed in two different detector elements, thus contributing to signal and noise;

3. it is scattered in the shielding and absorbed in a detector element, thus contributing to noise; or

4. it is absorbed in the shielding between the elements, contributing to neither noise nor signal.

There is thus an optimum in the detector shielding width between noise suppression and fill factor. We define a signal-to-noise ratio as

$$\text{SNR} = \frac{N_1}{\sqrt{N_1 + N_2 + N_3}}, \quad (5)$$

where $N$ is the number of photons with subscripts according to the three first cases above.

### 2.3.3. Separation of scattering from photoelectric events

Photons that are fully absorbed in a single detector element can be well characterized in terms of energy. Compton scattered photons, on the other hand, deposit energy in the detector corresponding to a probability distribution that depends on the photon energy (Eqs. 2 and 3). Because of the random component, Compton scattered photons provide very little energy information and it is important to be able to separate these from photoelectric events for energy resolved measurements. The separation may be accomplished using the deposited energy. Another alternative would be to use information about the depth of interaction, but we ruled out this method as inefficient prior to the present study.

## 3. RESULTS

Table 1 lists $N_\text{A}$, $N_\text{DM}$, and the total of these ($N_\text{tot}$) in 30 and 40 mm detectors for x-ray tube acceleration voltages 80-140 kVp. Also shown is the fraction of photoelectric interactions over Compton scattering (PE/C) at the different spectra. The detection performance is better for softer spectra, although for a 40 mm detector the difference between the spectra is not large. Softer spectra can, however, be expected to have a better energy resolution due to the larger amount of photoelectric interactions.

**Table 1.** The absorption efficiency ($N_\text{A}$), the transmission of the dicing margin ($N_\text{DM}$), and the total of these ($N_\text{tot}$) in 30 and 40 mm detectors for acceleration voltages 80-140 kVp. Also shown is the fraction of photoelectric interactions over Compton scattering (PE/C) at the different spectra.

| acc. voltage | 30 mm detector | | | | 40 mm detector | | | |
|---|---|---|---|---|---|---|---|---|
| [kVp] | PE/C | $N_\text{A}$ | $N_\text{DM}$ | $N_\text{tot}$ | PE/C | $N_\text{A}$ | $N_\text{DM}$ | $N_\text{tot}$ |
| 80 | 1.01 | 0.81 | 0.97 | 0.78 | 1.01 | 0.85 | 0.97 | 0.82 |
| 120 | 0.58 | 0.75 | 0.98 | 0.73 | 0.56 | 0.81 | 0.98 | 0.79 |
| 140 | 0.47 | 0.73 | 0.98 | 0.71 | 0.45 | 0.79 | 0.98 | 0.78 |

Figure 2 shows $N_\text{Th}$ as a function of noise threshold for 80-140 kVp spectra. Because of the fairly steep decrease, it is important to keep the noise threshold as low as possible. At 1 keV, $N_\text{Th} = 0.94$.

To determine the optimum shielding thickness, the SNR as a function of shielding thickness is plotted in Fig. 3 for a 30 mm detector element with a noise threshold of 1 keV. The optimum thickness is close to 20 $\mu$m for all considered tube voltages, which yields a fill factor $N_\text{F} = 0.96$.

The deposited energy spectrum in a 30 mm detector via scattering and the photoelectric effect is plotted in Fig. 4 for acceleration voltages 80, 120 and 140 kVp. At 80 kVp, the two types of events are well separated, but for 120 and 140 kVp there is a slight overlap.

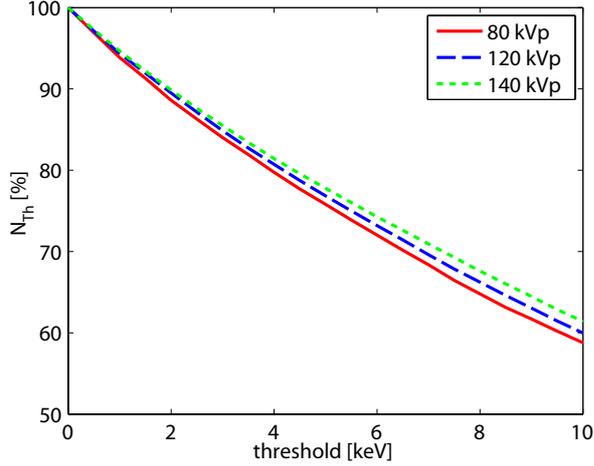
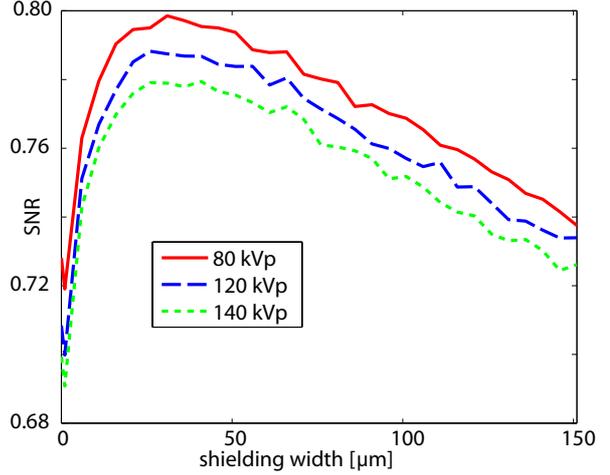

**Figure 2.** The detection efficiency as a function of lo-energy threshold ($N_{\text{Th}}$) for 80-140 kVp spectra.

**Figure 3.** The SNR as a function of shielding thickness.

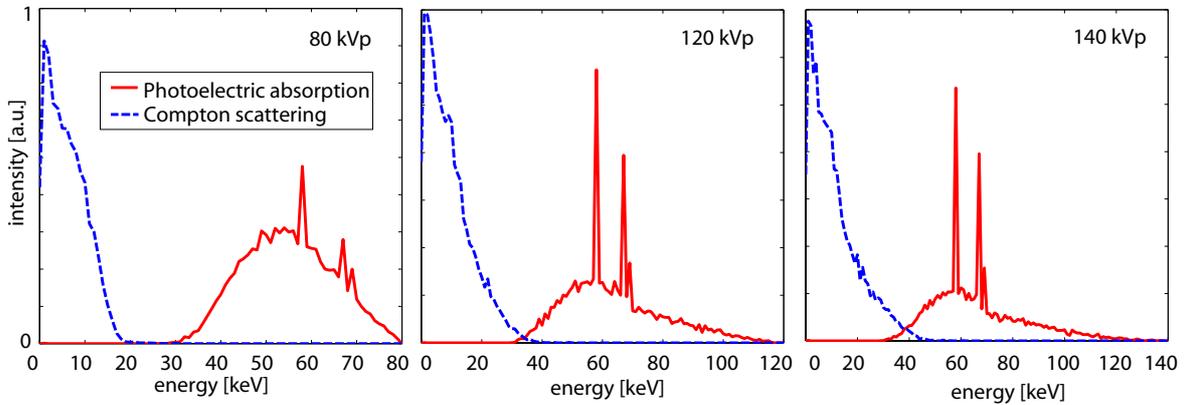

**Figure 4.** Deposited energy spectra in a 30 mm detector via scattering and the photoelectric effect for acceleration voltages 80 (left), 120 (middle) and 140 kVp (right).

## 4. DISCUSSION AND CONCLUSIONS

We propose to use the presented silicon detector with relatively soft spectra in order to maximize detection efficiency and minimize the fraction Compton scattered photons in the detector. Softer spectra might also yield a dose advantage in pediatric imaging.

A 20 μm tungsten foil between the detector units was found to be an optimal compromise between fill factor and scatter rejection in the detector. In order not to lose Compton events, the noise threshold needs to be minimized. With a 1 keV threshold, the detection efficiency of a 40 mm detector at 80 kVp is $N_{\text{A}} \cdot N_{\text{DM}} \cdot N_{\text{Th}} \cdot N_{\text{F}} = 0.74$. At this acceleration voltage, Compton scattered and photo absorbed photons are well separated in the energy spectrum.

In this study, a point source has been assumed, and a decline in detection efficiency can be expected for an extended and/or misaligned source because of increased absorption in the shielding. A positive effect of this angular sensitivity is, however, efficient intrinsic rejection of radiation that is scattered by the object. Also not included in the study is the expected degradation of the spatial resolution due to scattering between pixels in the same detector unit.

Photon-counting detectors are associated with a dead time, caused by the detector material and the readout electronics and leading to a decreased detection efficiency at high count rates. A major advantage of silicon is good charge diffusion properties, which minimizes the contribution of the detector material to dead time.

In conclusion, we have demonstrated that the shortcomings of silicon as a detector material can be largely overcome for pediatric CT at low acceleration voltages. A few issues remain to be investigated, including the angular sensitivity and pixelation of the detector, as well as the feasibility of readout electronics with low noise and dead time.

## ACKNOWLEDGMENTS

This study was funded by Familjen Erling-Perssons stiftelse.